\documentclass{PoS}

\usepackage{epsfig}

\newcommand{\gtapprox}{\raisebox{-0.5ex}{$\,\stackrel{>}{\scriptstyle\sim}\,$}}

\PoS{PoS(LAT2005)315}

\title{The pseudoparticle approach for solving path integrals in gauge theories}

\ShortTitle{The pseudoparticle approach for solving path integrals in gauge theories}

\author{
\speaker{Marc~Wagner} and Frieder~Lenz \\ 
Institute~for~Theoretical~Physics~III, University~of~Erlangen-N{\"u}rnberg, Staudtstra{\ss}e~7, 91058~Erlangen, Germany \\
E-mail: \email{mcwagner@theorie3.physik.uni-erlangen.de}, \email{flenz@theorie3.physik.uni-erlangen.de}
}

\abstract{
We present a numerical technique for calculating path integrals in non-compact U(1) and SU(2) gauge theories. The gauge fields are represented by a superposition of pseudoparticles of various types with their amplitudes and color orientations as degrees of freedom. Applied to Maxwell theory this technique results in a potential which is in excellent agreement with the Coulomb potential. For SU(2) Yang-Mills theory the same technique yields clear evidence of confinement. Varying the coupling constant exhibits the same scaling behavior for the string tension, the topological susceptibility and the critical temperature while their dimensionless ratios are similar to those obtained in lattice calculations.
}

\FullConference{
XXIIIrd International Symposium on Lattice Field Theory \\
25-30 July 2005 \\
Trinity College, Dublin, Ireland
}

\begin{document}


\section{Introduction and basic principle}

In this work we present a numerical method for calculating path integrals in Euclidean spacetime. We apply this method which we call pseudoparticle approach to U(1) and SU(2) gauge theories. It is a natural generalization of an idea that was successfully used to show the confining properties of ensembles of merons and regular gauge instantons \cite{LeNe04,NeLe04}.

In the following we explain the basic principle by concentrating on the case of SU(2) Yang-Mills theory\footnote{
The corresponding formulas for the U(1) case are very similar. Most of them arise by dropping color indices.
}. We perform Monte-Carlo calculations of SU(2) path integrals
\begin{eqnarray}
\Big\langle \mathcal{O} \Big\rangle & = & \frac{1}{Z} \int DA \, \mathcal{O}[A] e^{-S[A]} \\
S[A] & = & \frac{1}{4 g^2} \int d^4x \, F_{\mu \nu}^a(x) F_{\mu \nu}^a(x) \\
F_{\mu \nu}^a(x) & = & \partial_\mu A_\nu^a(x) - \partial_\nu A_\mu^a(x) + \epsilon^{a b c} A_\mu^b(x) A_\nu^c(x)
\end{eqnarray}
by representing the gauge field $A_\mu^a$ by a linear superposition of a fixed number $N$ of pseudoparticles:
\begin{eqnarray}
\label{EQN001} A_\mu^a(x) & = & \sum_{i=1}^N \rho^{a b}(i) a_\mu^b(x - z(i)) .
\end{eqnarray}
By a pseudoparticle we denote a field configuration $\rho^{a b}(i) a_\mu^b(x_\mu - z_\mu(i))$ which has localized action, energy and topological charge ($\rho^{a b}(i)$: amplitudes and color orientations; $z_\mu(i)$: positions). A common example is the regular gauge instanton, $a_\mu^b(x) = 2 \eta_{\mu \nu}^b x_\nu  / (x^2 + \lambda^2)$. The integration over all gauge fields is replaced by a finite dimensional integral over the amplitudes and color orientations of the pseudoparticles and can be computed by Metropolis sampling:
\begin{eqnarray}
\int DA \, \ldots & \rightarrow & \int \prod_{i=1}^N d\rho(i) \, \ldots .
\end{eqnarray}

For certain classes of pseudoparticles it can be shown that in the ``continuum limit'' (the limit of infinitely many pseudoparticles) every gauge field can be expanded as in (\ref{EQN001}) \cite{LeWa05}. That is in this limit the pseudoparticle approach is equivalent to the ``full'' SU(2) Yang-Mills theory. Therefore we expect to get a good approximation of SU(2) Yang-Mills theory even when using a finite number of properly chosen pseudoparticles ($\approx 400$ for the results in the following sections).

In our calculations Euclidean spacetime is restricted to a four-dimensional hypersphere. Inside the hypersphere the positions of the pseudoparticles are chosen randomly. Observables are measured sufficiently far away from the border.


\section{Maxwell theory}

To check our method we applied it to Maxwell theory. We used Gaussian localized pseudoparticles
\begin{eqnarray}
a_\mu^a(x) & = & \eta_{\mu \nu}^a x_\nu e^{-x^2 / (2 \lambda^2)}\end{eqnarray}
and antipseudoparticles
\begin{eqnarray}
\bar{a}_\mu^a(x) & = & \bar{\eta}_{\mu \nu}^a x_\nu e^{-x^2 / (2 \lambda^2)} \\
\end{eqnarray}
which form a basis of all transverse gauge fields in the continuum limit:
\begin{eqnarray}
A_\mu(x) & = & \sum_i \rho^a(i) a_\mu^a(x - z(i)) + \sum_j \bar{\rho}^a(j) \bar{a}_\mu^a(x - z(j))
\end{eqnarray}
($\eta_{\mu \nu}^a = \epsilon_{a \mu \nu} + \delta_{a \mu} \delta_{0 \nu} - \delta_{a \nu} \delta_{0 \mu}$ and $\bar{\eta}_{\mu \nu}^a = \epsilon_{a \mu \nu} - \delta_{a \mu} \delta_{0 \nu} + \delta_{a \nu} \delta_{0 \mu}$  are the t'Hooft- and anti-t'Hooft-symbol). There is no need to include longitudinal gauge fields because these fields neither appear in the field strength nor in the action. The integration over all gauge fields is replaced by an integration over the amplitudes of the pseudoparticles:
\begin{eqnarray}
\int DA \, \ldots & = & \int \left(\prod_{i,a} d\rho^a(i)\right) \left(\prod_{j,a} d\bar{\rho}^a(j)\right) \ldots .
\end{eqnarray}

We placed $250$ pseudoparticles and $250$ antipseudoparticles in a spacetime hypersphere of volume $500.0$. The average number density per unit volume in this ensemble is $n = 1.0$ and the average distance between neighboring pseudoparticles is $\bar{d} = n^{-1/4} = 1.0$. To be able to cover the whole spacetime hypersphere with pseudoparticles and to resolve as many ultraviolet details as possible the size of the pseudoparticles was set to $\lambda = \bar{d} = 1.0$.

\begin{figure}[b!]
\begin{center}
\input{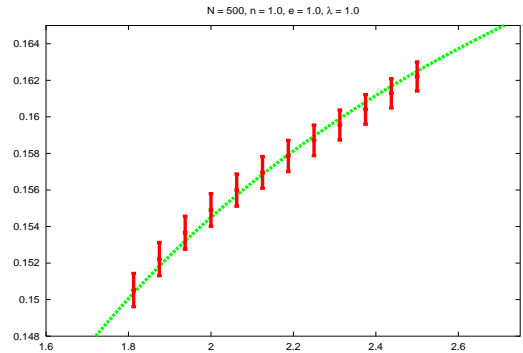}
\caption{\label{FIG001}Comparison of the potential of two static charges $\pm e$ as a function of their distance obtained with the pseudoparticle approach (red points with errorbars) and the Coulomb potential $V_{q \bar{q}}(R) = -e^2 / 4 \pi R$ (green curve).}
\end{center}
\end{figure}
The potential between two static quarks has been obtained by calculating rectangular Wilson loops and using a method \cite{Stac83} which is based on the well known formula
\begin{eqnarray}
V_{q \bar{q}}(R) T + C \approx -\ln \Big\langle W_{(R,T)} \Big\rangle .
\end{eqnarray}
The numerical results and the analytically known Coulomb potential $V_{q \bar{q}}(R) = -e^2 / 4 \pi R$ are plotted in Figure~\ref{FIG001}. For distances $R \gtapprox \lambda = \bar{d}$ they are in excellent agreement ($\lambda = \bar{d}$ determines the minimal size of the field fluctuations; it therefore acts as an ultraviolet cutoff and hence has a similar role as the lattice spacing in lattice calculations).


\section{SU(2) Yang-Mills theory}

For SU(2) Yang-Mills theory we used $1/r$-decreasing pseudoparticles because it has already been shown that pseudoparticles with long range interactions are a good choice of parameterizing the Yang-Mills path integral \cite{LeNe04,NeLe04}. We used three types of pseudoparticles:
\begin{eqnarray}
\label{EQN002} & & \hspace{-0.7cm} a_\mu^a(x) \ \ = \ \ \eta_{\mu \nu}^a \frac{x_\nu}{x^2 + \lambda^2} \quad \textrm{(``instanton'')} \\
\label{EQN003} & & \hspace{-0.7cm} \bar{a}_\mu^a(x) \ \ = \ \ \bar{\eta}_{\mu \nu}^a \frac{x_\nu}{x^2 + \lambda^2} \quad \textrm{(``antiinstanton'')} \\
\label{EQN004} & & \hspace{-0.7cm} \hat{a}_\mu^a(x) \ \ = \ \ \delta^{a 1} \frac{x_\mu}{x^2 + \lambda^2} \quad \textrm{(akyron)} .
\end{eqnarray}
(\ref{EQN002}) and (\ref{EQN003}) are similar to an instanton and an antiinstanton and the akyron\footnote{
Ancient Greek: \textit{akyros} $=$ pure gauge (literally ``without effect'').
} (\ref{EQN004}) is needed to get a complete set of functions in the continuum limit. As before the gauge field is a sum over pseudoparticles
\begin{eqnarray}
A_\mu^a(x) & = & \sum_i \rho^{a b}(i) a_\mu^b(x-z(i)) + \sum_j \bar{\rho}^{a b}(j) \bar{a}_\mu^b(x-z(j)) + \sum_k \hat{\rho}^{a b}(k) \hat{a}_\mu^b(x-z(k))
\end{eqnarray}
where $\rho^{a b}(i)$ is the product of the amplitude $\alpha(i)$ and the color orientation $u^{a b}(i)$ of the $i$-th pseudoparticle, i.\ e.\ $\rho^{a b}(i) = \alpha(i) u^{a b}(i)$. The color orientations are restricted by
\begin{eqnarray}
u^{a b}(i) & = & \delta^{a b} \Big(h_0(i)^2 - \mathbf{h}(i)^2\Big) + 2 h_a(i) h_b(i) + 2 \epsilon^{a b c} h_0(i) h_c(i)
\end{eqnarray}
with $h_0(i)^2 + \mathbf{h}(i)^2 = 1$. Analogous formulas hold for $\bar{\rho}^{a b}$ and $\hat{\rho}^{a b}$. The integration over all gauge fields is replaced by an integration over the amplitudes and color orientations of the pseudoparticles:
\vspace{-0.1cm} 
\begin{eqnarray}
\int DA \, \ldots & = & \int \left(\prod_i d\alpha(i) \, dh(i)\right) \left(\prod_j d\bar{\alpha}(j) \, d\bar{h}(j)\right) \left(\prod_k d\hat{\alpha}(k) \, d\hat{h}(k)\right) \ldots .
\end{eqnarray}
($dh(i)$, $ d\bar{h}(j)$ and $d\hat{h}(k)$ are invariant integration measures on $S^3$).

\begin{figure}[t!]
\begin{center}
\input{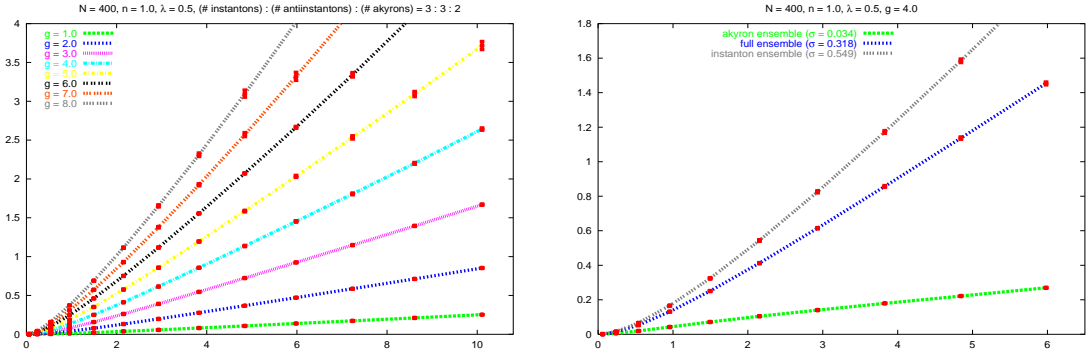}
\caption{\label{FIG002}
\textbf{a)} Negative logarithms of expectation values of rectangular Wilson loops $-\ln \langle W \rangle$ as a function of their area. Different curves correspond to different coupling constants.
\textbf{b)} Comparison of the ``akyron ensemble'' (green curve), the ``full ensemble'' (blue curve) and the ``instanton ensemble'' (grey curve).
}
\end{center}
\end{figure}

A crucial check of the pseudoparticle approach is to verify whether there is confinement or not. For that reason we calculated the expectation values of rectangular Wilson loops $\langle W_{(R,T)} \rangle$ with a fixed ratio of sides $R/T = 1/2$. We used $400$ pseudoparticles (ratio between ``instantons'', ``antiinstantons'' and akyrons $3:3:2$) with a number density per unit volume of $n = 1.0$ and a size of $\lambda = 0.5$. The results for different coupling constants are plotted in Figure~\ref{FIG002}a. The fact that there is an area law for large Wilson loops is a clear indication of confinement. We got numerical values for the string tension $\sigma$ by fitting an area perimeter function
\begin{eqnarray}
-\ln \Big\langle W_{(R,T)} \Big\rangle & \approx & a + b (R+T) + \sigma R T
\end{eqnarray}
in the usual way.

The pseudoparticle approach offers the possibility to characterize the importance of certain field configurations with regard to confinement. For instance we can turn off the instanton-like field configurations (equations (\ref{EQN002}) and (\ref{EQN003})) or the akyrons (equation (\ref{EQN004})). Figure~\ref{FIG002}b shows that the ``full ensemble'' and the ``instanton ensemble'' both exhibit area laws while in the ``akyron ensemble'' the perimeter term is dominating. Performing area perimeter fits yields $\sigma_\textrm{\scriptsize akyron} = 0.034$, $\sigma_\textrm{\scriptsize full} = 0.318$ and $\sigma_\textrm{\scriptsize instanton} = 0.549$, that is $\sigma_\textrm{\scriptsize akyron} : \sigma_\textrm{\scriptsize full} : \sigma_\textrm{\scriptsize instanton} \approx 1 : 9 : 16$. This together with the fact that a superposition of akyrons has vanishing topological charge density everywhere supports the expectation that confinement and topological charge are closely related.

\begin{figure}[b!]
\begin{center}
\input{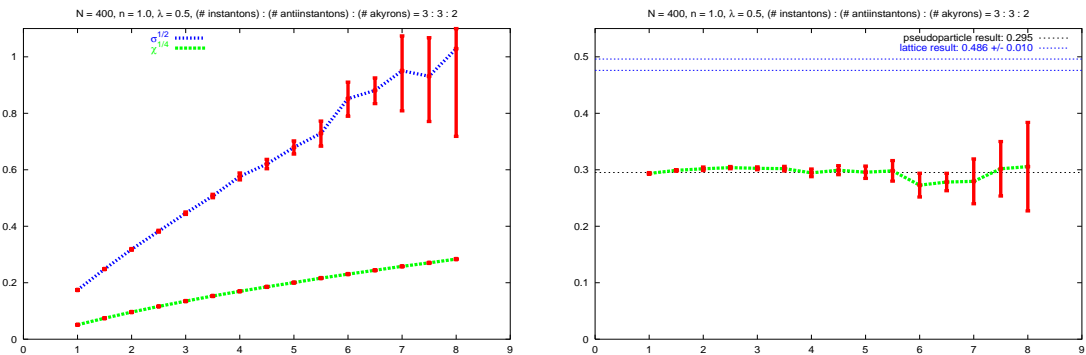}
\caption{\label{FIG003}
\textbf{a)} Dimensionful quantities $\chi^{1/4}$ (green curve) and $\sigma^{1/2}$ (blue curve) as functions of the coupling constant.
\textbf{b)} Dimensionless ratio $\chi^{1/4} / \sigma^{1/2}$ as a function of the coupling constant.
}
\end{center}
\end{figure}

For a quantitative comparison with lattice results we calculated the topological susceptibility $\chi = \langle Q^2 \rangle / V$. Figure~\ref{FIG003}a shows the variation of the dimensionful quantities $\chi^{1/4}$ and $\sigma^{1/2}$ with the coupling constant $g$. By setting the string tension to the physical value $\sigma = 4.2 / \textrm{fm}^2$ we can calculate the size of our spacetime hypersphere. For $1.0 \leq g \leq 8.0$ the diameter ranges between $0.51 \, \textrm{fm}$ and $2.91 \, \textrm{fm}$. In contrast to dimensionful quantities the dimensionless ratio $\chi^{1/4} / \sigma^{1/2}$ is nearly independent of the coupling constant (Figure~\ref{FIG003}b). Its value $0.295$ is in qualitative agreement with the lattice result $\chi_{\textrm{\scriptsize lattice}}^{1/4} / \sigma_{\textrm{\scriptsize lattice}}^{1/2} = 0.486 \pm 0.010$ \cite{Tepe98}.

\begin{figure}[t!]
\begin{center}
\input{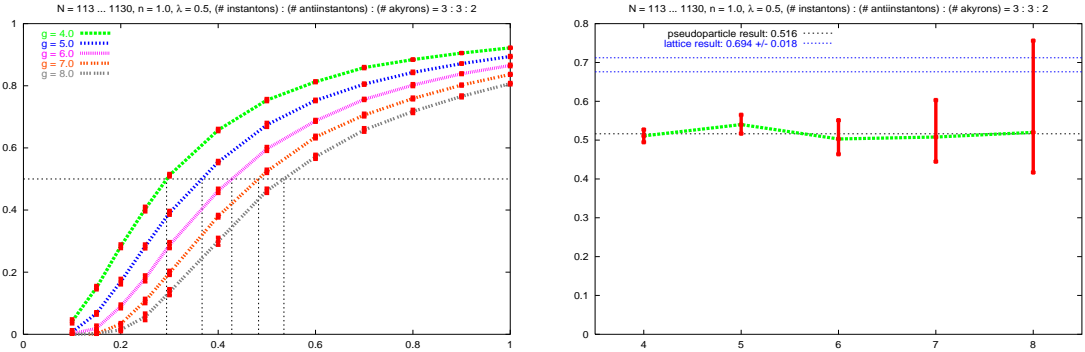}
\caption{\label{FIG004}
\textbf{a)} Expectation value of the Polyakov loop $\langle L \rangle$ as a function of the temperature. Different curves correspond to different coupling constants.
\textbf{b)} Dimensionless ratio $T_\textrm{\scriptsize critical} / \sigma^{1/2}$ as a function of the coupling constant.
}
\end{center}
\end{figure}

We also calculated the critical temperature of the confinement deconfinement phase transition. To this end we computed the temperature dependence of the expectation value of the Polyakov loop which serves as an order parameter. The results for different coupling constants are plotted in Figure~\ref{FIG004}a. Since we consider a finite system and since our ensemble is only approximately center symmetric we can not expect to see an exact phase transition. We define the ``critical temperature'' $T_\textrm{\scriptsize critical}$ to be that temperature where the expectation value of the Polyakov loop drops below a certain threshold, i.\ e.\ $\langle L \rangle(T_\textrm{\scriptsize critical}) = 1/2$. As shown in Figure~\ref{FIG004}b the dimensionless ratio $T_\textrm{\scriptsize critical} / \sigma^{1/2}$ is nearly independent of the coupling constant and its value $0.516$ is similar to the lattice value $T_{\textrm{\scriptsize critical, lattice}} / \sigma_{\textrm{\scriptsize lattice}}^{1/2} = 0.694 \pm 0.018$ \cite{Tepe98}.


\section{Conclusions}

We have shown that the pseudoparticle approach applied to Maxwell theory reproduces the Coulomb potential for static charges. At the same time the pseudoparticle approach yields an area law for Wilson loops in SU(2) Yang-Mills theory. Our results are in qualitative agreement with lattice data. We have investigated which properties of field configurations are responsible for confinement. Our findings support the common view that topological charge and confinement are closely related.



\end{document}